\documentclass[12pt,preprint]{emulateapj}
\usepackage{graphicx}
\usepackage{subfigure} 
\usepackage{natbib} 
\usepackage{float}
\usepackage{apjfonts}

\shorttitle{Progressive Red Shifts of SNe~Ia Spectra} 
\shortauthors{Black, Fesen, \& Parrent}

\begin{document}

\title{Progressive Red Shifts in the Late-Time Spectra of Type Ia Supernovae}

\author{C. S.\ Black\altaffilmark{1}, R. A.\ Fesen\altaffilmark{1}, 
       \& J. T. Parrent\altaffilmark{2}  } 
\affil{\altaffilmark{1}6127 Wilder Lab, Department of Physics \& Astronomy, Dartmouth College, Hanover, NH 03755 }
\affil{\altaffilmark{2}Harvard-Smithsonian Center for Astrophysics, 60 Garden St., Cambridge, MA 02138, USA}

\begin{abstract}

We examine the evolution of late-time, optical nebular features of Type Ia
supernovae (SNe~Ia) using a sample consisting of 160 spectra of 27 normal
SNe~Ia taken from the literature as well as unpublished spectra of SN
2008Q and ASASSN-14lp.  Particular attention was given to nebular features
between 4000$-$6000~\AA\ in terms of temporal changes in width and central
wavelength.  Analysis of the
prominent late-time 4700~\AA\ feature shows a progressive central wavelength
shift from $\sim$4600~\AA\ to longer wavelengths out to at least day +300 for
our entire sample.  We find no evidence for the feature's red-ward shift slowing or
halting at an [\ion{Fe}{3}] blend centroid $\sim$4700~\AA\ as has been proposed.
The width of this feature also steadily increases with a FWHM $\sim$170~\AA\ at day +100 growing to 200~\AA\ or more by day +350. Two weaker
adjacent features at around 4850 and 5000~\AA\ exhibit similar red shifts to
that of the 4700 \AA \ feature but show no change in width until very late
times.  We discuss possible causes for the observed red shifts of these 
late-time optical features including contribution from [\ion{Co}{2}]
emission at early nebular epochs and the emergence of additional features at
later times.  We conclude that the ubiquitous red shift of these common late-time SN Ia spectral features is not mainly due to a decrease in line velocities of forbidden Fe emissions, but the result of decreasing line velocities and opacity of permitted Fe absorption lines.

\end{abstract}


\section{Introduction}

Type Ia supernovae (SNe~Ia) have been instrumental in building our knowledge of
the Universe, from standard candles on cosmological scales to iron enrichment
of the interstellar medium (ISM) on a local scale.  SNe~Ia are thought to be the explosions of
degenerate carbon-oxygen white dwarfs that undergo a thermonuclear runaway when
they reach the Chandrasekhar limit \citep{HoyFow60,Nom84,Hil00}.
Although we currently do not have direct evidence of the progenitor
system(s) that leads to type Ia supernovae, a number of possible progenitors
have been proposed (see reviews by \citealt{How11,Nug11,BloMat12}). 

SN~Ia light curves are powered by the decay of $^{56}$Ni into $^{56}$Co
($t_{1/2}$ = 6.08 days) at early times, and then $^{56}$Co into $^{56}$Fe ($t_{1/2}$
= 77.2 days) beginning some two months after outburst. Near maximum light,
their spectra are characterized by strong permitted lines of intermediate-mass
and iron-peak elements \citep{Bra93}. SNe~Ia are believed to undergo a transition from a photospheric
phase to a nebular phase as the ejecta become transparent and the spectra
are increasingly shaped by iron-group emission features some 70$-$100
days past maximum brightness \citep{Bow97,Bra08,Fri14}.  

The homogeneity in the late-time optical spectra of normal SN~Ia suggest
a remarkably uniform and repeatable explosion model \citep{Bra08}.  While
photospheric phase spectra of SNe~Ia are numerous, so called nebular
spectra obtained some 70 days or more past maximum light are still fairly rare, with
less than 30 SNe~Ia with optical spectra past day~$+$200 \citep{Chi15}.
Consequently, the identification and evolution of some features in the nebular
phase have not been firmly established.

Differences in late-time feature identification exist in synthetic models
which match the observed late-time Ia spectra using: a) only
forbidden lines mainly from [\ion{Fe}{2}], [\ion{Fe}{3}], [\ion{Ni}{2}] and
[\ion{Co}{3}] \citep{Axe80,Bow97,StrMazSol06,Maz15,Chi15}, b) only permitted
lines, mostly from \ion{Fe}{2} and \ion{Cr}{2} \citep{Bra08,Bra09},
or c) both permitted and forbidden lines \citep{Fri14,Blo15}.
Uncertainty about the true nature of late-time spectral features limits
our understanding regarding such parameters as the relative abundances,
distribution and expansion velocities of Fe-peak element rich ejecta in SNe~Ia.

\begin{figure*}[t!]
        \centering
        \includegraphics [width=\linewidth]{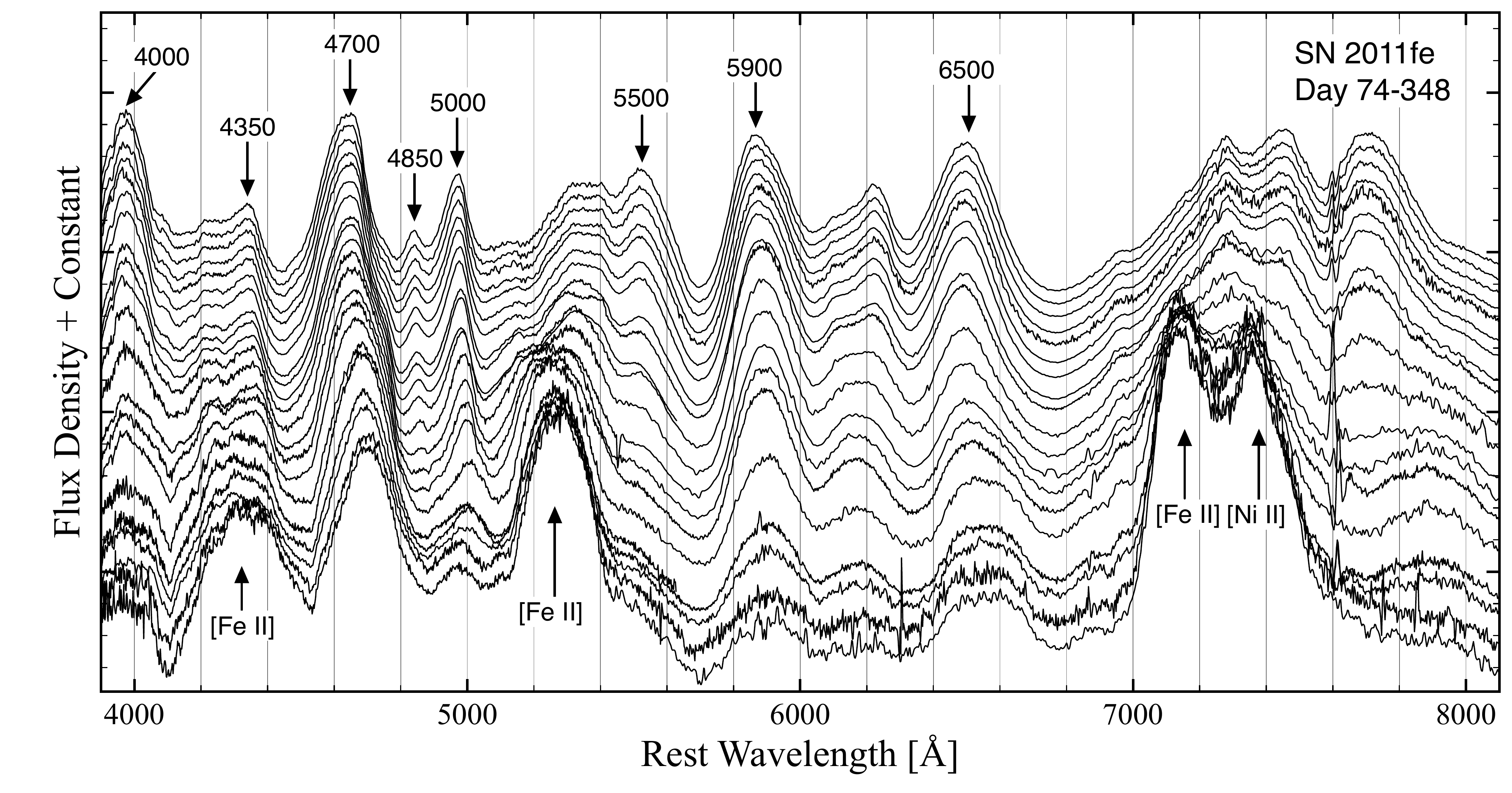}
        \caption{Rest-frame nebular spectra of SN 2011fe from $+$74 (Top) to $+$348 (Bottom) days 
                 are shown with some conventional late-time line identifications.}
        \label{fig:11fe}
\end{figure*}

The identification of even the most prominent feature in optical late-time
SN~Ia spectra between 4600 and 4700~\AA, which dominates the optical and IR regions, is controversial.  \citet{Mae11} and \citet{Sil13} report
finding a gradual shift toward longer wavelengths of this feature in many
late-time SNe~Ia spectra which they interpret as a blueshift of $-4000$ to $-2000$ km
s$^{-1}$ blueshift of a 4701~\AA \ blend of [\ion{Fe}{3}] emission lines which
decreases to $\sim$0 km s$^{-1}$ when the feature red shifts to $\simeq$4700
\AA \ by day +300. 

However, an [\ion{Fe}{3}] blend interpretation has difficulty explaining the
position of the feature at very late-times (day $>$ 300) when it can appear red-ward
of  4700~\AA\ \citep{Pan15}. In addition, the progressive red
shifting of other late-time SN~Ia spectral features first reported
by \citet{Min39} has not been explained or adequately addressed in recent late-time SN~Ia
studies.  Instead of a Doppler shift of forbidden line emission, \cite{Bra09}
argue that the 4700~\AA\ feature may be photospheric emission shaped by permitted P-Cygni line
absorptions that appear shortly after maximum and persist well into what is conventionally viewed as the  nebular phase.

Here we present the results of an investigation of late-time optical spectra of
27 normal SNe~Ia at multiple epochs with the goal of better understanding the
evolution and nature of late-time spectral features.  Our data set and
observations are described in \S2, results on the changes of central
wavelengths for certain late-time features described in \S3, with our
discussion of these results given in \S4. In \S5 we discuss several computer
model results concerning late-time, blue spectral features of SNe~Ia, and
summarize our findings and conclusions in \S6.

\section{Data Set} \label{sec:Data}

The centerpiece of our comparative analysis is the normal SN~Ia~2011fe, whose
post-maximum light phases were extensively observed between days $+$74 and
$+$348.  Our analysis of late-time optical spectral properties includes
a compilation of rest-frame nebular SN~Ia spectra taken from the literature,
on-line databases, and a few new observations described below.  

Several publicly available SN databases were used to develop a sample of
high-quality, late-time Ia spectra.  These databases include the on-line
Supernova Spectrum archive (SUSPECT; \citealt{Ric01}), the Weizmann Interactive
Supernova Data Repository (WISeREP; \citealt{Yar12}), and the UC Berkeley
Supernova Database (SNDB) which is a part of the Berkeley SN~Ia Program, or
BSNIP \citep{Sil12}.

Our archival sample of data contains 155 optical,
late-time spectra of 26 ostensibly normal SNe~Ia observed between
days $+$60 and $+$400 from SUSPECT, WISeREP, and BISNP, and have been corrected for
redshift.  Some of the SNe~Ia examined here are also included in the late-time
samples of \citet{Mae10Jan}, \citet{BloMat12}, and \citet{Sil13}.

\subsection{SN 2008Q}\label{sec:08Qobs}

Our SN~Ia sample includes three nebular observations of SN~2008Q which occurred
in NGC~524 (z~=~0.00802).  A 1200s spectrum was taken on 28 August 2008 (day
+203)  with the 10m Keck telescope using the Low Resolution Imaging
Spectrometer (LRIS). This spectrum was published in the study by \citet{Sil13}.
Two additional 1200s exposures of SN~2008Q presented here were taken on 30 September and 20
December 2008 (days $+$237 and $+$318) with the MMT 6.5m using the Blue Channel
Spectrograph with the 300 lines mm$^{-1}$ grating.

\subsection{ASASSN-14lp}\label{sec:14lpobs}

SN~Ia ASASSN-14lp in NGC 4666 (z = 0.00510) was the second brightest SN of
2014, reaching a peak magnitude of V~=~11.94 mag on 24 December 2014 with a
$\Delta$m$_{15}$ of 0.79 \citep{Sha15}.  Moderate resolution spectra (R $\sim$
2000$-$3000) of ASASSN-14lp were taken with exposure times of 2400s and 4000s
in March and April 2015  at days $+$87, $+$114, respectively, with the MDM 2.4m
using the Ohio State Multi-Object Spectrograph (OSMOS).  A follow-up 900s
spectrum was obtained in May 2015 (day~$+$149) with the South African Large
Telescope (SALT) using the Robert Stobie Spectrograph (RSS).  Data reduction
was done using \textsc{iraf}\footnote{\textsc{IRAF} is distributed by the National Optical
Astronomy Observatories, which are operated by the Association of Universities
for Research in Astronomy, Inc., under cooperative agreement with the National
Science Foundation.} and consisted of bias and background subtraction,
wavelength calibration, aperture extraction, and host galaxy redshift
correction.

\begin{figure}[t!]
    \centering
    \includegraphics[width=\columnwidth]{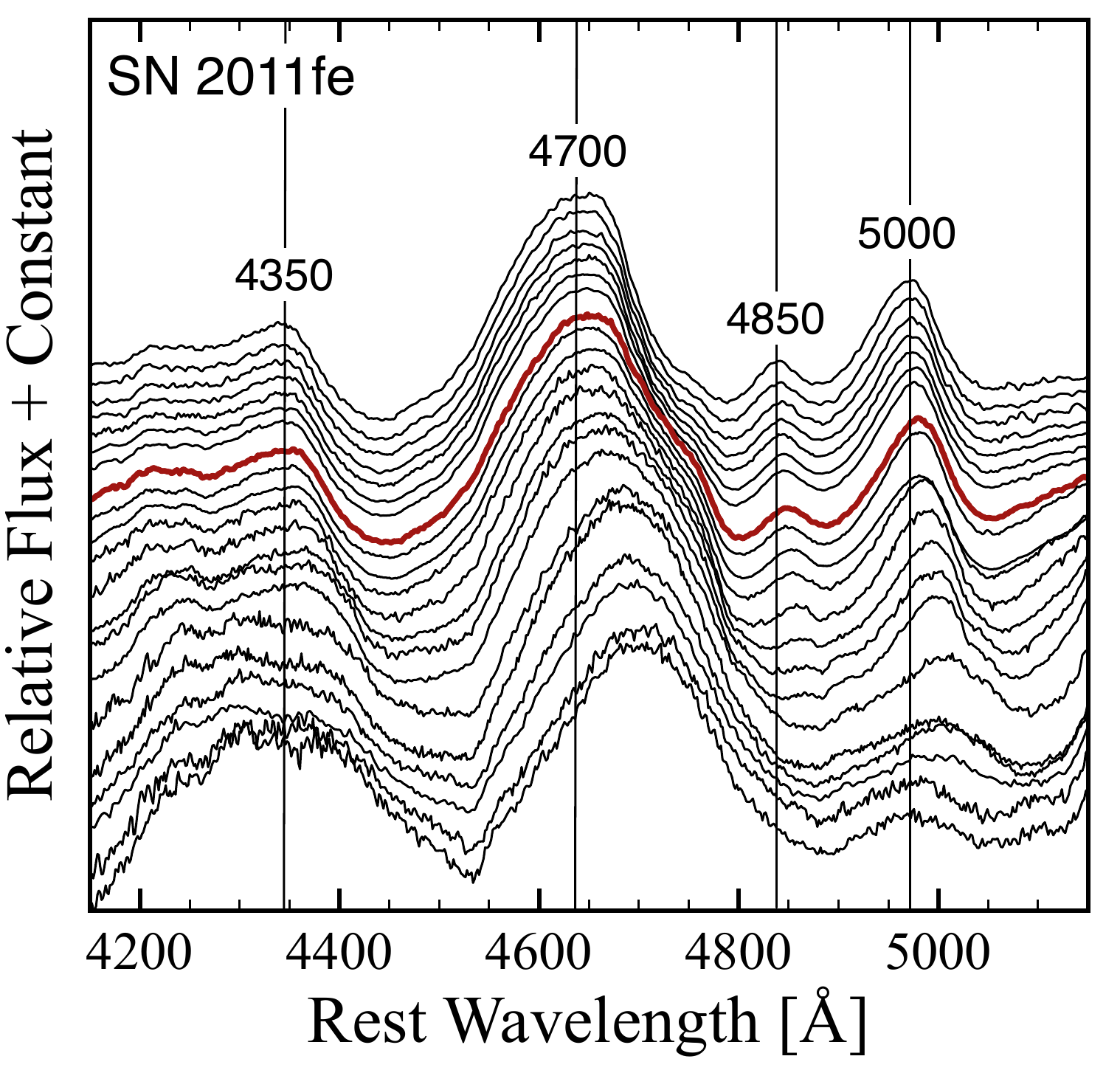}
    \caption{Post-maximum light spectra of SN 2011fe between 4200 and 5100~\AA\ 
             for days +74$-$348. The solid lines are centered on the 4350~\AA, 4700~\AA, 
             4850~\AA, 5000~\AA\ features at day~$+$74, highlighting the shift 
             to redder wavelengths exhibited by some features.  
             Bold red line marks day +100.}
    \label{fig:474850}
\end{figure}

\begin{figure*}[htb!]
        \centering
           \includegraphics[width=\textwidth] {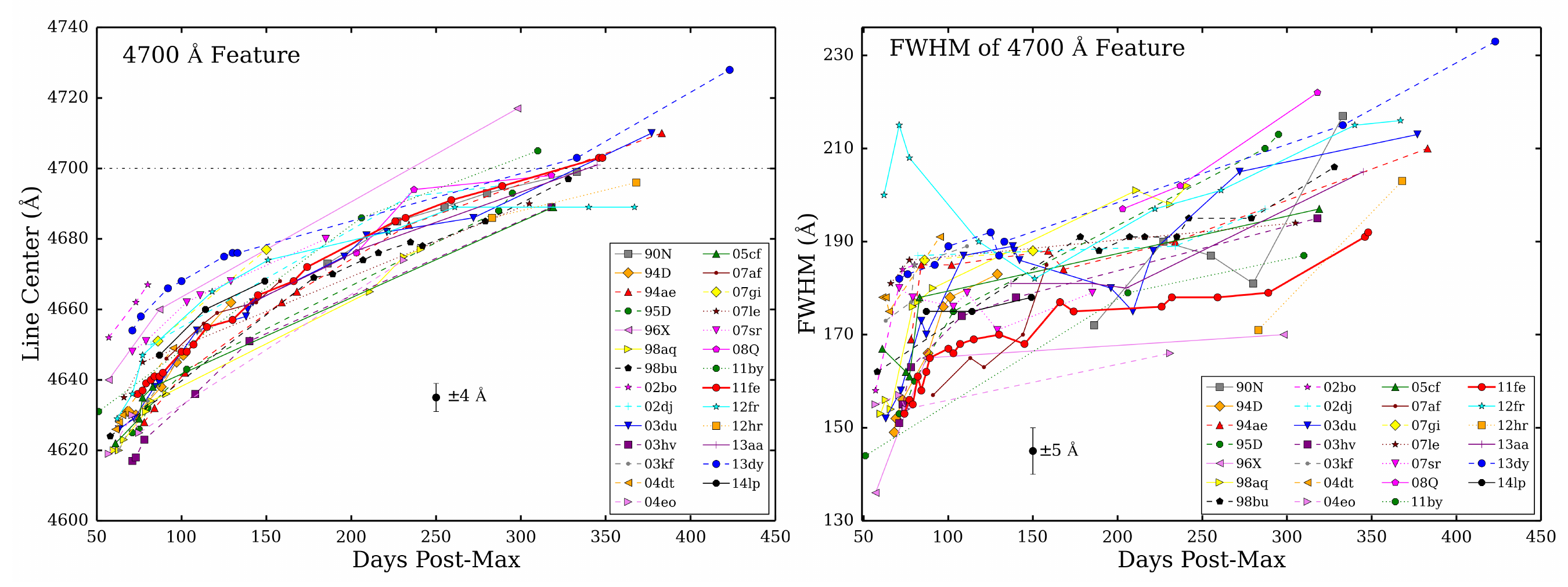}
        \caption{{\it Left}: Central wavelength of the 4700~\AA\ feature for all SN
                 in this study versus days post-maximum light. This feature appears to follow the same
                 evolution for all normal SNe~Ia.  The gray line represents the rate of change exhibited by the 		4700~\AA\ feature.  The horizontal black dot-dashed line marks 4700~\AA.
                 {\it Right}: Evolution of the FWHM of the 4700~\AA\ feature.}
        \label{fig:4700_allfwhm}
\end{figure*}

\begin{figure*}[t!]
    \centering
    \includegraphics[width=0.7\textwidth]{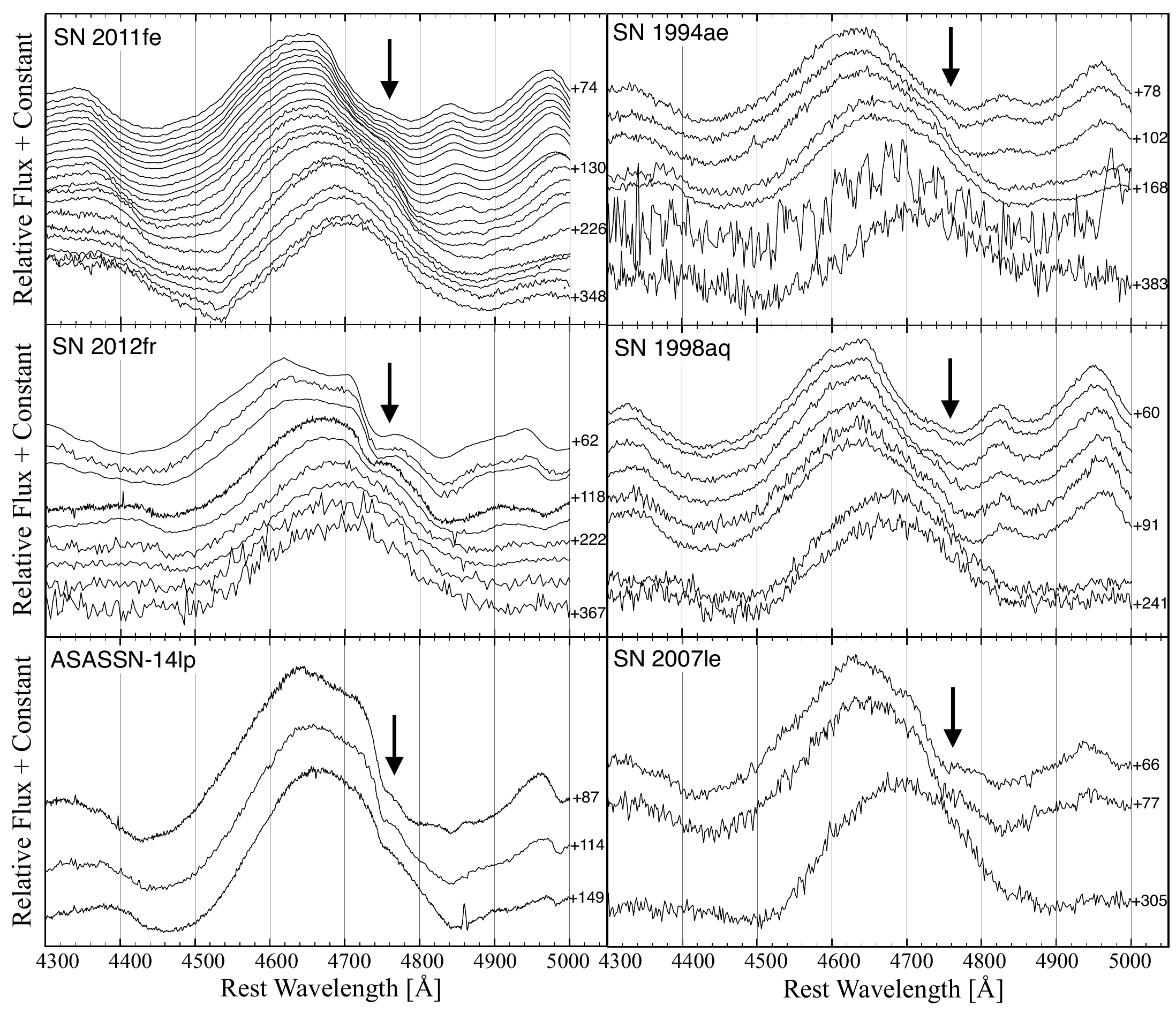}
    \caption{Several SN~Ia show evidence of additional flux on the red side of the 4700~\AA\ feature.  {\it Top}: SN 2011fe, SN 1994ae; {\it Middle}: SN 2012fr, SN 1998aq;
             {\it Bottom}: ASASSN-14lp, SN 2007le}
    \label{fig:blend}
\end{figure*}

\section{Results}\label{sec:Move}

The general evolution of late-time optical SNe~Ia spectra is illustrated in the
data for SN~2011fe shown in Figure~\ref{fig:11fe}.  SN~2011fe was well
observed at late-times, making it a good case to
investigate the late-time spectral evolution of a normal SNe~Ia.

As can be seen in this figure, significant changes occur from the early nebular
phase at day $+$74 out to late-times at day~$+$348 \citep{Mat08,Par14}.  Some
features present at early epochs weaken and fade, while other prominent
features do not emerge until much later times.  

Below we explore the evolution of a few selected features in the SN~2011fe and
other late-time SN~Ia spectra, focusing on the 4000 to 6000~\AA \ region since
it contains many of the brightest late-time optical features.  We begin by
investigating apparent red shifts of several late-time, blue spectral features.

\subsection{Progressive Red Shifts of Features}\label{redshifts}

Progressive red shifts in the central wavelengths of four late-time features
marked as 4350, 4700, 4850, and 5000 during the nebular phase of the spectra of
SN~2011fe can be seen in Figure~\ref{fig:474850}.  Vertical lines mark the
central wavelengths of these features at day $+$74. The 4350~\AA\ feature is
severely blended making it difficult to accurately measure changes in its
central wavelength at late times.  Thus, this feature is not included in this
study.  The other three features can be seen to not only shift toward the red
but also move at seemingly similar rates. However, at around day~$+$100 (marked
in red) the red side of the 4700 \AA\ feature shows the emergence of some additional
emission which appears to play a role in red shifting this feature's central
wavelength.  We discuss below the detailed properties and wavelength changes of each of
these three features.

\subsubsection{The 4700~\AA \ Feature}

One of the strongest late-time, optical spectral SN~Ia features is seen between
4600 and 4700~\AA \ and has most often been referred to as the 4700~\AA\ feature.  It is commonly identified as a blend of [\ion{Fe}{3}] lines at 4658, 4702,
4734~\AA\ centered near 4700~\AA \ (\citealt{Bow97,Maz15,Chi15}, and references therein). However,
there may be additional contributions from weaker [\ion{Fe}{3}] emissions at
4755, 4769, and 4778~\AA\ \citep{Mae10Jan}, [\ion{Fe}{2}] 4814~\AA\ \citep{Maz15},
as well as [\ion{Co}{2}]~4623~\AA\ \citep{Liu97Nov} 

Measured line centers of the 4700 \AA\ feature for the 27 SNe~Ia in our sample
show a systematic shift to longer wavelengths with time. This is shown in the left panel of
Figure~\ref{fig:4700_allfwhm}.  In this work, the central
wavelengths were measured with an estimated error for low S/N spectra of
$\pm$4~\AA\ by fitting a Gaussian to the feature using the Spectral Analysis
Tool (SPLAT; \citealt{Dra14}).  A table listing the measured line centers for each supernova can be found in the Appendix.

In the spectra of SN~2011fe, indicated by the thick red line in Figure 3,
the feature appears centered around 4635~\AA\ at day~$+$74, shifting steadily
towards $\sim$4700~\AA\ by day~$+$348.  All 26 other SNe~Ia in our sample show
a similar red shift and rate of wavelength change.  

Although the shift is continuous, it is not strictly linear with time.  In the
early nebular phase of SN~2011fe between days $+$50 and $+$125, the 4700 \AA\
feature shifts $\sim$0.3~\AA\ per day (20~km~s$^{-1}$ per day), slowing down to
a nearly constant rate of $\sim$0.2~\AA\ per day (12~km~s$^{-1}$ per day), in
agreement with the findings of \citet{Mae11} and \citet{Sil13}.  This
transition from 0.3 to 0.2 \AA\ per day is not abrupt but occurs
gradually during days 100-125.  Although few SNe in our sample have spectra past $\sim$300 days,
we find no indication that the late-time rate of $\sim$0.2~\AA\ per day
redshift changes past this date.

Along with a progressive red shift, the width of the 4700~\AA\ feature is
also seen to increase with time.  The broadening of this feature in the spectrum of SN~2011fe can be readily seen in
Figure~\ref{fig:474850}.  At around day~$+$100, the red side shows the
apparent emergence of some additional emission, which is not seen in the other neighboring features.  This could explain the steady
increase of the 4700~\AA\ feature's FWHM after day $+$100.  

Measured FWHM values of the 4700~\AA\ feature for all 27 SNe are plotted in the
right hand panel of Figure~\ref{fig:4700_allfwhm}.  The FWHM was measured using
SPLAT with respect to the base of the emission feature as it rises above
adjacent continuum with an estimated error of $\pm$5~\AA.  
Our values are some 20\% to 30\% smaller than reported values by
\citet{Maz98}, \citet{BloMat12}, and \citet{Sil13}, which is likely due to differences in chosen 
adjacent flux levels. 

For all SNe in our sample, the FWHM of the 4700~\AA\ feature appears to
increase relatively rapidly between days $+$50 and $+$100, going from
$\sim$150 to $\sim$175~\AA\ ($\sim$0.5 \AA\ per day).  This increase coincides with the feature's
rapid change in central wavelength.  At later times, the FWHM continues to
increase but at a slower rate of $\sim$0.1~\AA\ per day out to day~$+$400.  

The widening of the 4700~\AA\ feature caused by the emergence of added flux at day $+$100 may
not be uncommon.  Examination of other late-time SN~Ia spectra in our sample
revealed five other SNe exhibiting strong line blending on the red side of the
4700~\AA\ line profile much like that seen in SN~2011fe. 

In Figure~\ref{fig:blend}, we show the region around the 4700~\AA\ feature for
SN~1994ae, SN~1998aq, SN~2007le, SN~2012fr, and ASASSN-14lp. Arrows mark the
same rest wavelength of the additional flux as seen in SN~2011fe.  The profile of the 4700~\AA\ feature is
strongly asymmetric for several of these SNe before the appearance of the
additional red-ward emission.  This is especially clear for SN~2012fr and
ASASSN-14lp, and significantly influences the FWHM of this feature for SN
2012fr prior to day $+$100 (see Fig.~\ref{fig:4700_allfwhm}).  The emergence
of this flux seems to occur earlier in some objects, around day +60 in
some cases, compared to SN~2011fe.

\subsubsection{The 4850 and 5000~\AA \ Features}

\begin{figure*}[t!]
        \centering
           \includegraphics[width=\textwidth] {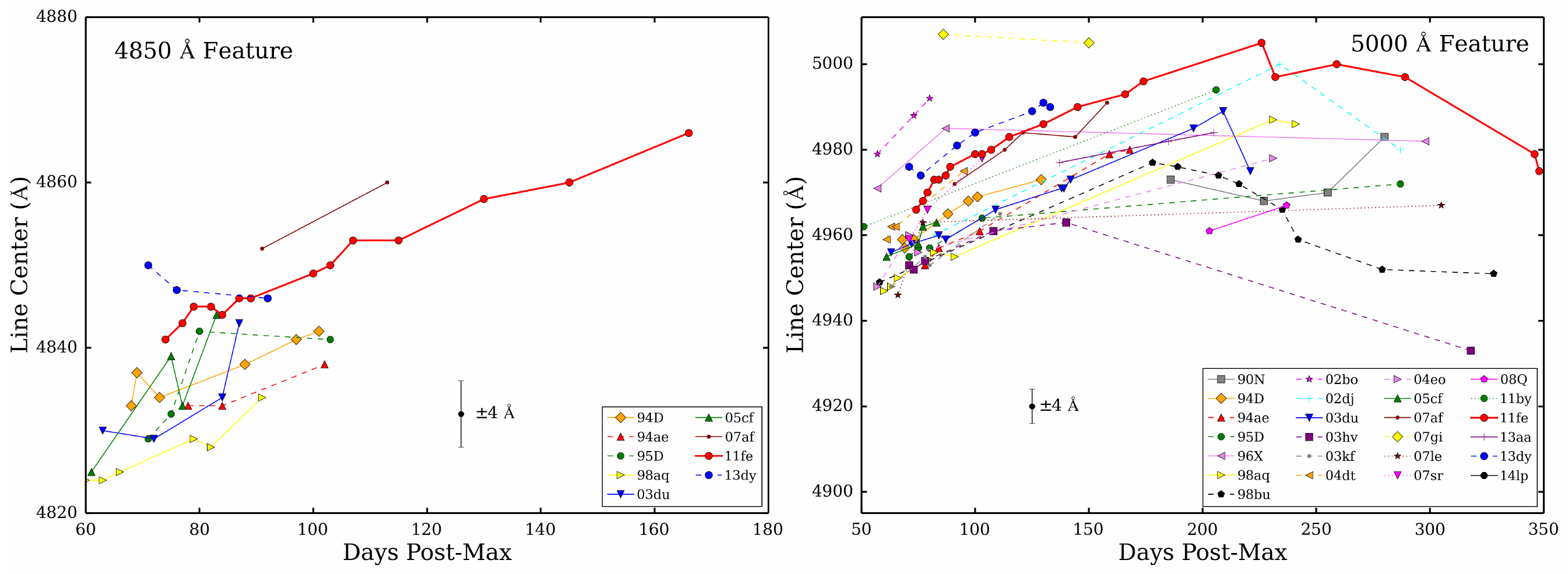}
        \caption{{\it Left}: Central wavelength of the 4850~\AA\ feature for all SNe 
               	in this study versus days post-maximum light.
	 	{\it Right}: Central wavelength of the 5000~\AA\ feature for all SNe
                 in this study versus days post-maximum light.
                 The gray line represents the rate of change exhibited by the 4700~\AA\ feature.}
        \label{fig:4800_5000}
\end{figure*}

\begin{figure}
    \centering
    \includegraphics[width=\columnwidth]{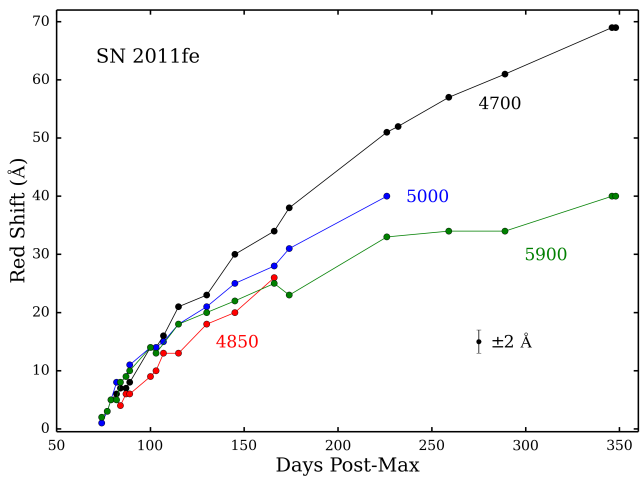}
    \caption{The central wavelengths of the 4700, 4850, 5000, and 5900 \AA\ features
of SN 2011fe comparing the relative red shift rates.}
    \label{fig:11ferate}
\end{figure}

In normal late-time SNe~Ia spectra, two relatively weak features appear red-ward
of the 4700~\AA\ feature, around 4850 and 5000~\AA\ (see Fig.~\ref{fig:11fe}).
Both emerge as early as day +7 \citep{Mat08,BloMat12,Par14}. The 4850~\AA\
feature typically fades by around day +75 to +100, in contrast to the feature at 5000~\AA\
which is a long-lived feature that can be seen out to day $\sim$+1000 \citep{Tau15,Gra15}.
Although the 4850~\AA\ feature has not been previously identified, the stronger feature at $\sim$5000~\AA\ has
been identified as [\ion{Fe}{2}] \citep{Maz11} or [\ion{Fe}{3}] \citep{Maz15}
or a combination of both \citep{Liu97Nov,Bow97,Pas07}.

Similar to that seen for the 4700~\AA\ feature, both the 4850 and 5000~\AA\ features move
progressively toward longer wavelengths with time, as shown in
Figure~\ref{fig:474850}.  The 4850~\AA\ feature shifts roughly ~0.3~\AA\ per day
from day $\sim$$+$50 to $+$100, much like the early red shift seen for the 4700~\AA\
feature.  This red shift can be seen in the left hand panel
of Figure~\ref{fig:4800_5000} where in SN 2011fe the feature moves roughly 20~\AA\
between days $+$74 and $+$140. The long lived visibility of the 4850~\AA\ feature in
SN~2011fe seems to be an exception.  

The 5000~\AA\ feature likewise exhibits a substantial redshift where its line
center shifts some 40~\AA; from $\sim$4960~\AA\ at day $+$74 to $\sim$5000~\AA\ at day
$+$200.  The motion of this feature in 25 out of the 27 SN~Ia from our sample is shown in the right hand panel of Figure~\ref{fig:4800_5000}.
Although some SNe~Ia were better observed at late-times than
others, we estimate a wavelength shift $\sim$~0.25~\AA\ per day during from day
$+$100 to $+$200.

After about day~$+$200, the 5000~\AA\ feature appears to stop moving red-ward
and, in some cases, broadens with a line center shift toward shorter wavelengths.  This
apparent motion could signal the presence of weak emission visible only at very
late times.  The feature's FWHM remains fairly constant during the early
nebular phase with a width of $\sim$75~\AA\ ($+$70 days) but then increases to
$\sim$100~\AA\ at late times ($+$200 days).

The three 4700, 4850, and 5000 \AA\ features appear to drift towards longer
wavelengths at virtually the same rate between days $+$50 to $+$150.  This is
shown in Figure \ref{fig:11ferate} where the central wavelength shifts are
overlain.  While the red shift of these features are nearly identical at
early times, the rate of red shift of the 4700~\AA\ feature is relatively larger at later
times, likely due to the appearance/disappearance of minor blended features after day $+$100,
as mentioned above.

Finally, unlike the 4700~\AA\ feature, the peaks of the 4850 and 5000~\AA\ features are
relatively narrow and do not show any obvious indication of line blending.  This suggests that these features
are showing a truer evolution of the line velocities compared to the 4700~\AA\
feature. See \S4.4 where we discuss the possibility that they are pseudo-emission features due to P-Cygni Fe II absorption and that the observed shift originates from decreasing line velocity and opacity of Fe II.

\subsection{Emission at 5900~\AA \ }

\begin{figure*}[t!]
        \centering
           \includegraphics[width=\textwidth] {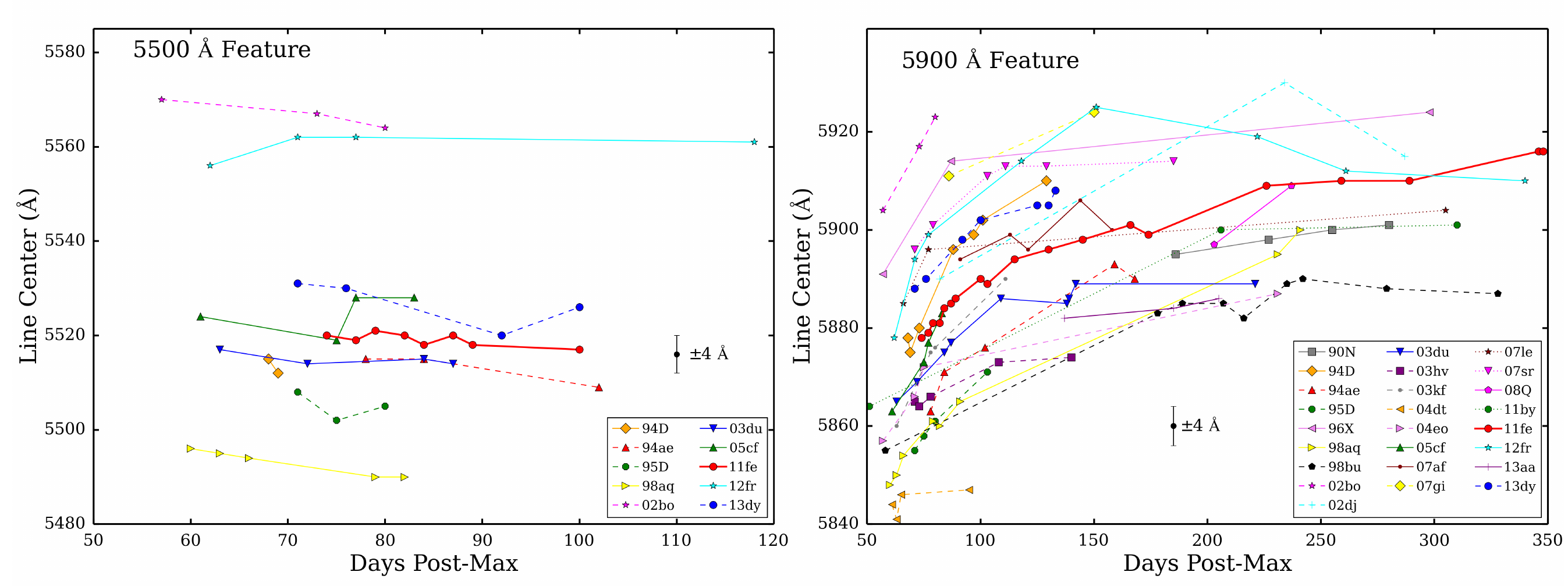}
        \caption{ {\it Left}: Central wavelength of the 5500~\AA\ feature for all SNe 
               	in this study versus days post-maximum light.
	 	{\it Right}: Central wavelength of the 5900~\AA\ feature for all SNe
                 in this study versus days post-maximum light.}
        \label{fig:5900}
\end{figure*}

Like the emission features between 4350 and 5100~\AA, the strong late-time emission
feature around 5900~\AA\ also shifts to the red but at a significantly different
rate.  The 5900~\AA\ feature shows a rapid change in wavelength of 0.4 $\pm0.2$~\AA\ per day from early nebular epochs to about day~$+$100. At later times, this
rate decreases to $\sim$0.1~\AA\ or less per day, just half the rate of the 4700~\AA\ blend.
This difference can be seen in SN~2011fe, where its red shift is nearly identical to the 4700 and
5000~\AA\ features but then slows down significantly past day 150 (see Fig.~\ref{fig:11ferate}).

For many SNe in our sample,
including SN~2011fe, the line center of the 5900~\AA\ feature redshifts beyond
5890~\AA, which is redward of the wavelengths of both [\ion{Co}{3}] 5888~\AA\ and
\ion{Na}{1} 5890, 5896~\AA. In addition, the 5900~\AA\ feature exhibits a larger range of line center wavelengths at a given epoch (see Figure~\ref{fig:5900}).

Whereas the evolution of the
4700~\AA\ feature spans roughly 20~\AA\ for our sample, the 5900~\AA\ feature shows
a range of approximately 50~\AA.  Finally we note that there may be a weak correlation between the range in central wavelengths of the 5900 \AA\ feature and the red/blueshifts seen in the [\ion{Fe}{2}] 7155 \AA\ and [\ion{Ni}{2}] 7378 \AA\ features noted by \citet{Mae10Jan,Mae10July,Mae11}.

The 5900~\AA\ feature has been conventionally attributed to a blend of [\ion{Co}{3}]
5888, 5907~\AA\ along with possibly weak \ion{Na}{1} 5890, 5896~\AA \ emission, and
recent models suggest that \ion{Na}{1} may not significantly shape this feature
at all \citep{DesHil14,Chi15}.  However, unlike the strong 4700~\AA\ feature which disappears at very late times
(day $>$ +500; \citealt{Tau15,Gra15}), the 5900~\AA\ feature remains present even at
very late times (day +980; \citealt{Gra15}) appearing at $\sim$5880~\AA\, blue-ward of
its center wavelength location of 5910~\AA \ at day +350.  Since there is
little [\ion{Co}{3}] expected at such late times, consistent with the decline
of [\ion{Co}{3}] emission at 6600 \AA, some of the observed emission might
be due to \ion{Na}{1} 5890, 5896~\AA\ emission \citep{Gra15}.

\section{Discussion}\label{sec:disc}

Several weeks after maximum, the optical spectra of SNe~Ia mainly consist of strong features around 4000, 4600, 5900, 7400 and 8700~\AA. The
4000~\AA\ feature has been attributed to either P-Cygni \ion{Ca}{2} H~\&~K emission
and/or [\ion{S}{2}] 4072~\AA \ while most of the other features have typically been
attributed to forbidden lines of [\ion{Fe}{2}], [\ion{Fe}{3}], [\ion{Co}{2}], and [\ion{Co}{3}]
\citep{Bow97,Mae10Jan,Sil13,Maz15}.  One year after maximum brightness, strong [\ion{Fe}{2}] line blends
at 4370, 5270, 7155, and 8617~\AA \ plus emission from [\ion{Ni}{2}] 7378~\AA\
appear in the optical spectrum (see Fig.~\ref{fig:11fe}). 

Nebular SN~Ia spectra have most often been interpreted as mainly
composed of forbidden lines of iron and cobalt \citep{Mey78,Mey80,Axe80,Bra90}
and early spectral modeling supported this idea with models that generated
moderately good fits to observed spectra lines (e.g., SN~1991T;
\citealt{Spy92}). Late-time spectral studies have shown the
importance of optical and near-IR [\ion{Co}{2}] and [\ion{Co}{3}] lines and the
dominance of singly and doubly ionized Fe in late-time SN~Ia spectra up to at
least day $\sim$300 \citep{Bow97,Pen14,Chi15}.

However, because optical spectra of SNe~Ia suffer from severe line blending,
obtaining robust line profile measurements and accurate line identifications of
some features has been difficult. This is true even at
very late times and has led to studies focusing on the much
fainter near-infrared (NIR) lines of [\ion{Fe}{2}] and [\ion{Co}{2}], which are
more isolated and relatively unblended even when highly Doppler broadened.

The difficulty of determining firm line identifications for strong line blended,
late-time optical features of SNe~Ia has led to significant differences among several synthetic spectral models (e.g.,
\citealt{Bow97,Blo15,Bra09,Fri14}). This, in turn, has naturally led to some uncertainty in
the true composition of late-time optical SN~Ia spectra. Furthermore, the lack
of an explanation for the underlying cause for the large and progressive
wavelength shifts of some late-time features seen in SN~Ia spectra
discussed above only adds to the problem of securing robust feature
identification.  Below we discuss the implications of the results of our study
of late-time optical features in SN~Ia spectra.

\subsection{Previous Post-Maximum Red Shift Interpretations}

Recent investigations of the nebular phase spectra of SNe~Ia has revealed that
certain features exhibit significant but fixed blue or red rest wavelength
offsets \citep{Maz98, Mot06, Ger07, Mae10Jan, Mae10July, Maz15}.  This includes the
late-time optical emission lines [\ion{Fe}{2}] 7155~\AA \ and [\ion{Ni}{2}]
7378~\AA\ (see Fig.~\ref{fig:11fe}).  Such steady displacements of
[\ion{Fe}{2}] and [\ion{Ni}{2}] emissions have been interpreted as Doppler
shifts of up to $\pm$2000 km s$^{-1}$ resulting from differences in viewing
angles stemming from asymmetrical iron peak element rich ejecta in the center
of the expanding SN debris \citep{Mae11}.

However, viewing angle cannot explain the progressive and universal line
center red shifts seen for the 4700, 4850, and 5000~\AA\ features in our SN~Ia
sample as discussed above. If the observed shifts were due to viewing angle, one would expect that all
features in the late-time spectra would show red or blue shifts, a phenomena that is
not observed. 

Large red shifts for the 4700, 4850, and 5000~\AA\ features in
post-maximum SN~Ia was first noted and described by Minkowski in a report
on the late-time spectral evolution of the very bright and now recognized
prototypical Ia event, SN~1937C in IC~4182, along with SN~1937D in NGC~1003 and
two other SNe.  He noted that below 5000~\AA \ late-time SN~Ia spectrum
consists of several bands which shift ``as a whole, gradually toward the red
without any significant changes in its intensity distribution'' \citep{Min39}.
He also noticed a shift of the 5900~\AA\ feature shift, though not to the
extent shown here. 

Minkowski further noted that the red shifts of four features at 4350, 4600,
4850, and 5000~\AA \ to be large and virtually identical in the spectra of his
four best studied SNe.\footnote{A pseudo 3D graphical representation showing
Minkowski's measured wavelengths of emission features with respect to
post-maximum phase for SN~1937C in IC~4182 using shaded bands to indicate changes in
intensity can be found in \citet{Zwicky65}.} Measured from maximum light, the
especially broad and strong 4600~\AA \ feature seen in SN~1937C 
moved from 4575~\AA \ to around 4700 \AA \ between day 0 and +339, amounting to
a wavelength shift of some 125~\AA.

Subsequent late-time nebular spectra of other SNe~Ia made by a number of other
observers confirmed Minkowski's claim of progressive red shifting of features
between 4000 and 5000~\AA \
\citep{Ros60,vdBergh61,Ber62,Ber65,Chi64,Kik71,Kir73,Bar82}.  However, a detailed
analysis of late-time spectra of the especially bright SN~Ia event 1972E in
NGC~5253 seem to show only the 4600~\AA\ feature to drift in wavelength, with all
other features fixed in wavelength \citep{Kir73}.

Minkowski found that the size and rate of wavelength shifts of the four blue
spectral features was so strikingly similar at a given post-maximum phase that
he believed wavelengths measurements of these features could be used to
accurately estimate when a supernova had reached its maximum brightness. This notion was actually
employed by several subsequent observers for SNe Ia with poorly observed light
curves. 

Interestingly, few studies published between 1985 and 1995 on late-time SNe~Ia
spectra remark on the varying central wavelength of these blue spectral
features. In fact, many late-time spectral studies make no mention of prior
observations concerning progressive red shifting of any features. For example,
in their analysis of fixed Doppler displacements of the [\ion{Fe}{2}] and
[\ion{Ni}{2}] features in late-time spectra interpreted as due to view angle
effects, \citet{Mae10Jan} state that the 4700~\AA\ feature showed no Doppler shift
away from an assumed [\ion{Fe}{3}] blend at 4701~\AA.  

In subsequent papers, however, \citet{Mae11} reported a
large red shift of the 4700~\AA\ feature which they interpreted as a large initial
blueshift velocity.  They argued that in early nebular phases, the 4700~\AA \
[\ion{Fe}{3}] blend exhibits a blueshifted of some $-4000$ to $-2000$ km
s$^{-1}$ in SNe~Ia which then steadily decreases to $\sim$0 km s$^{-1}$ by day
+200.  \citet{Sil13} reported similar findings, noting the decrease in the
assumed blueshift velocity was both slow and approximately linear with time,
amounting to 10 to 20 km s$^{-1}$ per day. In a late-time study of SN~2013dy, \citet{Pan15} reported the [\ion{Fe}{3}]
4700 \AA\ emission blend to be redshifted by $+$1300 km s$^{-1}$ at day
+423. None of these studies offered an explanation for either the
large initial blueshift of the 4700~\AA\ feature or its change over to a redshift at
very late times.

\subsection{The Nature of the Progressive Red Shifts}

\citet{Min39} wrote at length concerning possible causes for the progressive and
uniform post-max red shifts of blue spectral features he saw in the late-time
spectra of SN~1937C and other type I supernovae. Although most red spectral
features underwent significant changes in intensity with time, he noted that
they showed no systematic change in their wavelengths like that seen in
the blue. He concluded that a Doppler or gravitation effect was possibly
responsible but cautioned that ``no attempt at identification of the bands in the
blue can give conclusive results so long as an explanation for the red shift is
lacking''.

\citet{Pay40} viewed the red shift as likely the result of increased ionization
of the expanding material with time. \citet{Min41} disagreed.  \citet{McL63}
puzzled over the nearly uniform patten of red shifting features and wondered if
it was simply the result of decreasing blueshifted absorption features on a
background continuum, but gave no details.  While \citet{Mus71} agreed that a
steady red shift could result from the gradual deceleration of absorption
lines, he argued that if true such a similar shift would then be expected for
red features, contrary to the observations.  He noted that the spectrum below 5000
\AA \ likely contains highly complex blends, making the drawing of
any kind of simple conclusions hazardous.  

\citet{Mey78} suggested that the red shifts was simply a transparency effect (see also \citealt{Fri12}).
That is, during the early post-maximum epochs emission lines such as from
[\ion{Fe}{3}] would only be observable from the approaching expanding
hemisphere resulting in a blueshifted Doppler line which would be seen to
decrease with time, in accord with observations.

\subsubsection{Red Shifting of the 4700~\AA\ Feature  }

As the strongest feature in the nebular phase, the 4700~\AA\ feature naturally attracted
special attention in many studies of late-time SN~Ia spectra. Below we address
the nature of the progressive red shifted feature by first discussing the properties
of this feature.

Interpreting it as due to [\ion{Fe}{3}] emission, \citet{Kuc94}
used it to calibrate the decay timescale of [\ion{Co}{3}] 5888, 5907~\AA\ by means of the
4600/5900 flux ratio.  Later, \citet{Maz98} and \citet{BloMat12} suggested its
FWHM was anti-correlated with $\Delta m_{15}$(B).  However, this has been questioned
by \citet{Maz12} and \citet{Sil13}, and our investigation shows that
the FWHM of the 4700 features varies with time, further undermining such a
correlation.

For SN~2011fe, the red shift of the 4700~\AA\ feature is substantial, amounting to
60~\AA \ between days +70 to +350.  As already noted above, while this
feature has been widely attributed to mainly a blend of 
[\ion{Fe}{3}] emission lines, there have been 
other identifications proposed.

In an attempt to model the late-time spectrum of the type Ia event SN~1994D,
\citet{Liu97July} found generally good agreement concerning the 4700~\AA\ feature which they viewed as chiefly due
to a blend of [\ion{Fe}{3}] emission.  However, because its observed strength was much
stronger than the predicted spectrum before day +200, this led them to suspect
other emissions might be contributing to this feature's strength in early
nebular phases.  

Consequently, \citet{Liu97Nov} proposed that the 4700~\AA\ feature contained a
significant contribution from radioactive Co on the feature's blue side due to
[\ion{Co}{2}] 4623~\AA\ emission. They suggested that as this Co line emission
decayed with time, the central wavelength of the 4700~\AA\ blend loses a strong
contribution at bluer wavelengths, effectively pushing the blend of Co and Fe emission lines
red-ward as observed. 

In this way, their emission model can help explain the initial rapid 0.3~\AA\ per day shift of the 4700~\AA\ feature that is observed.  The decreasing contribution from [\ion{Co}{2}]~4623~\AA, coupled with strengthening [\ion{Fe}{3}]~4658~\AA, will cause the feature to move to redder wavelengths more rapidly than compared to the other features until [\ion{Co}{2}] emission no longer dominates the blue side of the 4700~\AA\ feature.

\citet{Mae10Jan} noted that the 4700~\AA\ feature showed no Doppler shift away from
the presumed [\ion{Fe}{3}] blend at 4701~\AA\ and that the observed change in
wavelength is caused by either blending due to \ion{Mg}{1}] 4571~\AA\ or a
``radiation transfer effect". They argued that the absence of a Doppler shift
of the 4700~\AA\ feature, in contrast to the displacement of the [\ion{Fe}{2}]
7155~\AA\ line emission, was consistent with a picture where outer, low density and
higher ionization Fe$^{++}$ zone in SNe~Ia was more spherically symmetric. 

Although attributing the 4700~\AA\ feature to a blend of the $a^{5}$D$_{4,3,2}$ --
$a^{3}$F$_{4,3,2}$ 4658, 4702, and 4734~\AA \ [\ion{Fe}{3}] lines would seem
quite likely given the strong presence of [\ion{Fe}{2}] in very late-time
nebular spectra, identifying the feature as mainly due to [\ion{Fe}{3}] is not
without difficulty.  The strongest optical [\ion{Fe}{3}] feature is the
4658~\AA\ emission line and is always observed stronger than the combined
strength of the [\ion{Fe}{3}] lines 4702~\AA\ and 4734~\AA\ lines in both
photoionzied \ion{H}{2} regions \citep{Est04, Est13} and in the shock heated
environment of supernova remnants \citep{Fes96}, in agreement with theoretical
predictions \citep{Kee92,Kee01,DebHib09}.  It is therefore surprising that the
feature does not appear more asymmetric toward the blue at late-times. 

Moreover, we find no evidence that the rate of change in the feature's central
wavelength is slowing down or stopping at 4700~\AA \ in our 27 SN sample.
Instead, the feature appears at wavelengths greater than 4700~\AA\ for some
SNe~Ia at very late-times, e.g., SN 2013dy at 4728~\AA\ at $+$423 days (see
Fig.  \ref{fig:4700_allfwhm}). At such late-times, the supposed  [\ion{Fe}{3}]
blend is then centered more than 50~\AA\ away from the strong  [\ion{Fe}{3}]
4658 \AA\ line.

This does not mean that [\ion{Fe}{3}] is absent from this feature.  Late-time nebular spectra contains strong emission from [\ion{Fe}{2}] (see Fig.~\ref{fig:11fe}), so it is reasonable for some [\ion{Fe}{3}] emission to also be present.  However, it appears that [\ion{Fe}{3}] is not the main source for the 4700 \AA\ feature's intensity and central wavelength.

It is clear from our investigation that some of the observed redshifting of the
4700~\AA\ feature may also be due to the emergence of additional emission on the red
side starting around day $+$100.  This added emission, which may not be
uncommon (see Fig.\ \ref{fig:blend}), may also be responsible for the evolution
of the 4700~\AA\ feature's width where it steadily increases.  Additional
[\ion{Fe}{3}] emission lines at 4755, 4769, and 4778~\AA\ may be responsible
for some of this emission as suggested by \citet{Mae10Jan}, but these relatively
weak lines and would be expected to appear simultaneously with other
[\ion{Fe}{3}] lines.

\begin{figure*}[t!]
        \centering
        \includegraphics[width=\textwidth]{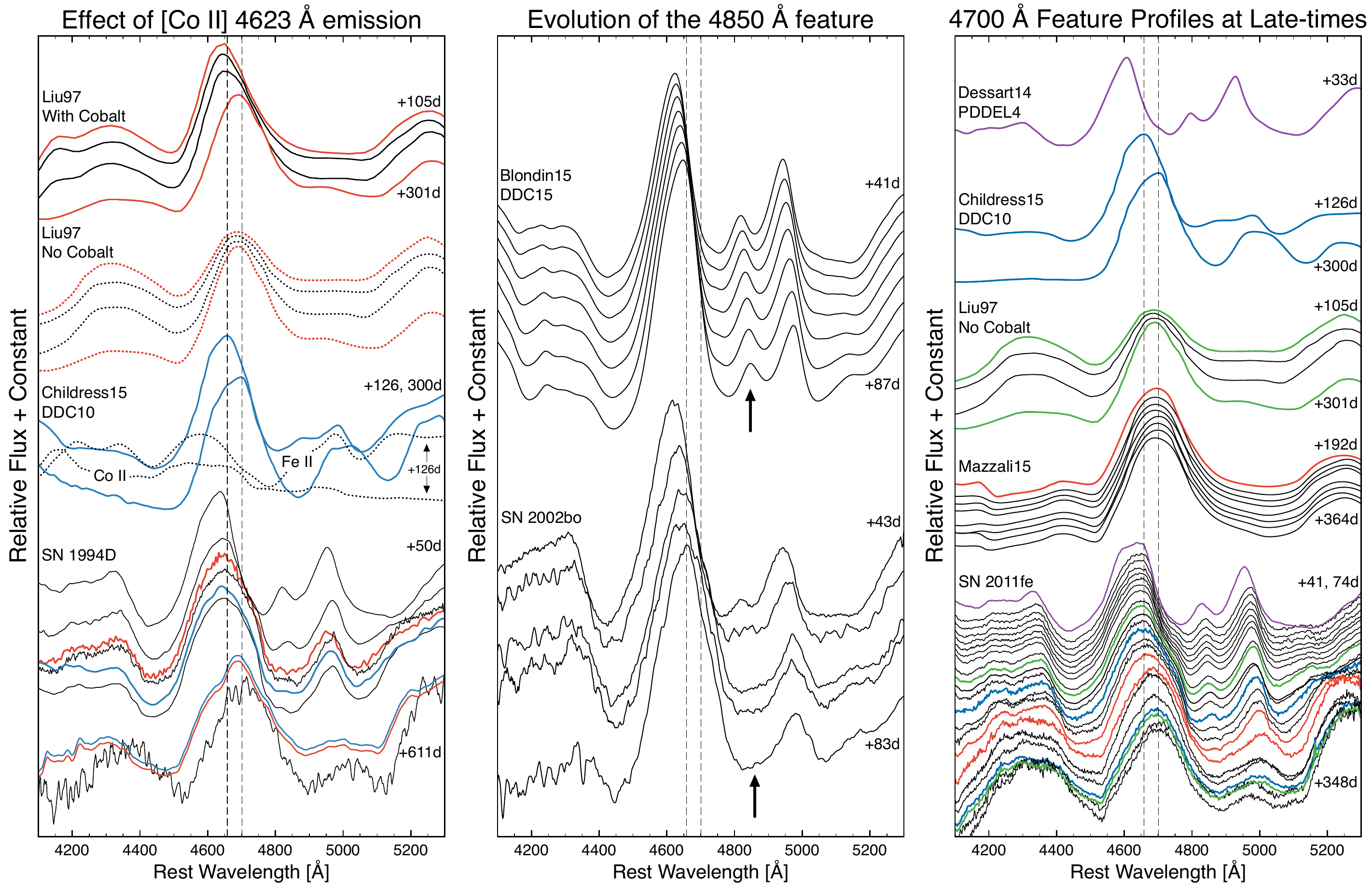}
        \caption{ {\it Left}: Late-time spectra of SN 1994D compared to the model spectra from \citealt{Liu97Nov} with and without Co emission, and the day~$+$126, 300 spectra for the DDC10 model discussed by \citet{Chi15}. {\it Center}: Late-time spectra of SN 2002bo compared to the DDC15 model presented by \citealt{Blo15}. {\it Right}: Late-time spectra of SN 2011fe compared to: \citet{Liu97Nov} without Co emission; the latest, available post-maximum spectrum from the
PDDEL4 model of \citealt{DesBlo14}; late-time spectra of DDC10 from \citet{Chi15}; and the late-nebular series produced by \citealt{Maz15}. The day~$+$611 spectrum of SN~1994D \citep{Pat96,Gom98,BloMat12} and the latest two observations of SN~2002bo \citep{Ben04,BloMat12,Sil12} have been
smoothed with a boxcar filter. Colors denote nearest epochs between a model and
the corresponding observations. Dashed-vertical lines are centered at
4658~and~4701~\AA, and arrows point to the feature near 4850~\AA. All spectra have been scaled with respect to the 4700~\AA\ feature. }
 \label{fig:models}
\end{figure*}

\subsubsection{Red Shifts of Other Nebular Features}

Whatever the nature of the red-ward motion for the 4700~\AA\ feature, neither the
decay of [\ion{Co}{2}] 4623~\AA\ emission or a decreasing blueshift of a blend of [\ion{Fe}{3}] line
emission can easily explain the origin of the similar redshifting for neighboring
features, namely those at 4850 and 5000~\AA.  Despite their near universal
presence in late-time SNe~Ia spectra, the 4850~\AA\ feature is usually 
left unidentified, while the 5000~\AA\ has been attributed to
[\ion{Fe}{2}] and/or [\ion{Fe}{3}].

Although the 5900~\AA\ feature also shifts towards longer wavelengths with time, it
does so at a different rate than the blue spectral features at late times.
After day~$+$100 the emission drifts red-ward at half the rate of the
4700, 4850, and 5000~\AA\ features.  This suggests the cause of its red shift
may be different than that seen in the blue.  In addition, the central
wavelength of the 5900~\AA\ feature varies significantly across our sample of
SN~Ia, spanning 50~\AA\ versus the 20~\AA\ exhibited by the 4700~\AA\ feature
at the beginning of the nebular phase. 

A slower red shift evolution might support the notion that this feature is
primarily due to [\ion{Co}{3}]~5888~\AA\ \citep{DesHil14}.   However, the line
center evolves red-ward of 5888~\AA\ after day~$+$200, possibly
indicating either the emergence of additional line emission, or an effect of a transparent, continuum-emitting core  \citep{Fri12}.  

It is interesting to note that some features that have been attributed to
forbidden cobalt emission show no significant red shifts.  For example, early
nebular epochs show a strong emission at 5550~\AA\ feature possibly from
[\ion{Co}{2}] 5544~\AA\ \citep{Liu97Nov} which at late times blends with the
broader 5300~\AA\ feature as it decays. Prior to blending with the neighboring
5300~\AA\ feature, the 5550~\AA\ emission region shows little change in wavelength, as shown in the left panel of Figure \ref{fig:5900}.
However, if the 5550 and 5900~\AA\ features are both chiefly due to forbidden Co
emission, it is strange that the 5900~\AA\ feature drifts while the 5550~\AA\ feature
does not.

\subsection{Progressive Red Shifts in Models of Late-Time SN~Ia Spectra}

Synthetic models provide a powerful avenue of investigation into late-time
spectra of SN~Ia, and in particular the phenomenon of the red-ward drifting
features.  Realistic models must be able to account for the movement of some
features, the lack of movement of others, as well as include weak features
like those of 4850 and 5000~\AA, all of which is no easy task.

In Figure~\ref{fig:models}, late-time spectra of three SN~Ia are plotted
covering the 4100 to 5300~\AA\ region and a sample of accompanying models.
These SN include: SN 1994D \citep{Pat96,Gom98,BloMat12}, SN 2002bo
\citep{Ben04,BloMat12,Sil12}, and SN 2011fe \citep{Per13,Maz14,Maz15}.

The five models shown in this figure include: a sub-M$_{ch}$ model with and
without Co emission computed by \citet{Liu97Nov} for SN~1994D; the
delayed-detonation model, DDC10, on days~$+$126 and $+$300, which were
discussed in \citet{Chi15} in comparison to SN~2011fe and 2012fr (see also
\citealt{Blo13}); the delayed-detonation model, DDC15, from day~$+$53 to
day~$+$104, which was favored by \citet{Blo15} for the broad-lined
SN~Ia~2002bo; the day~$+$50 spectrum of the pulsational-delayed detonation
(PDD) model, PDDEL4, which was favored by \citet{DesBlo14} for SN~2011fe; and
the late-nebular series produced by \citet{Maz15} from a M$_{ch}$-mass model
for SN~2011fe.  Spectra and models at similar epochs have been highlighted in
Figure~\ref{fig:models} for clarity. 

While the emission between 4600 and 4700 \AA\ is the strongest optical feature at
late times, its
progressive line center motion, increasing width, emergence of
blended emission around day~$+$100, and strong blending make this feature particularly difficult to model.  Furthermore, the models shown in Figure~\ref{fig:models} each have different
prescriptions regarding the rate and reasons for the evolution of the 4700~\AA\ feature.

The models computed by \citet{Liu97Nov} (left panel), \citet{Blo15} (center
panel), and \citet{Chi15} (right panel) appear to capture the observed
evolution of the 4700~\AA\ feature as it moves red-ward  with rates of roughly 0.23, 0.42, and 0.20 \AA\ per day, respectively.  Although, while these
models show red shifts similar to that actually observed, the origins of the shift differΠin terms of spectral decomposition. 

For the Liu et al.\
model, the red-ward drift is
primarily caused by both decreasing emission from [\ion{Co}{2}]~4623~\AA\ and
increasing emission from lines of [\ion{Fe}{3}].  This appears to be the case for the 4700 \AA\ feature for both the Blondin et al.\ and Childress et al.\ models, in addition to minor contributions from permitted and/or forbidden \ion{Fe}{2} emission. However, these models fail to reproduce the evolution of 4850 and 5000 \AA\ features, which presumably stem from signatures of iron.

In contrast, the
computed spectra of \citet{Maz15}, where the 4700~\AA\ feature is dominated by [\ion{Fe}{3}] emission
at all epochs, does not show the redward progression from $\sim$4658~\AA\ to 4701~\AA.
Instead, the peak of the 4700~\AA\ feature oscillates up to 10~\AA\ about
4701~\AA.

Curiously, compared to the broad peak of the 4700~\AA\ feature for SN
2011fe on day~$+$41, the same region for the PDDEL4 model does not appear broad
enough, suggesting an absence of unknown emission (see right panel of
Fig.~\ref{fig:models}). Similarly, the peak of the 4700~\AA\ feature for the
DDC10 model on day~$+$300 is asymmetric and leans toward 4701~\AA. This exact shape
is not observed for SN 2011fe during this phase, but this emission structure is
possibly seen for SN 1994D. 

Nevertheless, Figure~\ref{fig:models} reveals that models without a significant
contribution from P-Cygni \ion{Fe}{2} and [\ion{Co}{2}] 4623~\AA\ emission show no drifting,
while models that include contribution from \ion{Fe}{2} and [\ion{Co}{2}] not only predict a red-ward shifting 4700~\AA\ feature, but appear to
shift at the same rate as observed. Thus, an association between an [Fe III] emission blend at 4701~\AA\ and the peak would appear weak. Given also the lack of significantly
blue-shifted lines of Fe emission in the computed spectra of \citet{Liu97Nov}
between days $+$105 and $+$301, we find the 4700~\AA\ feature cannot be used to
infer evolving line-velocities of Fe.

As noted above, the minor features near 4850~\AA\ and 5000~\AA\ are generally
not accurately predicted by late-time models.  Of the models shown, only those
of \citet{DesBlo14} and \citet{Blo15} reasonably produce these features (see
center and right panels of Fig.~\ref{fig:models}). Spectra for the
corresponding PDDEL4 model have only been computed out to day~$+$33, but we
suspect the evolution of the 4850~\AA\ is similar to that of the DDC15 model
\citep{DesBlo14,Blo15}. 

The DDC15 model includes the constant red-ward drift seen in the 4700, 4850, and
5000~\AA\ features.  However, despite its prominence in this model, the
4850~\AA\ feature, indicated by the arrows in Figure~\ref{fig:models} is not
strongly observed in SN~2002bo. Other broad-lined SNe~Ia similarly reveal an
apparent lack of this 4850~\AA\ feature roughly two months since maximum light
(see Figs. 8$-$11 of \citealt{Par14}). Instead, the model from \citet{Blo15}
predicts that these features shift in time with little change in relative line
strengths between the 4850 and 5000~\AA\ features.  

\subsection{Permitted Lines at Late-times?}

While synthetic spectra built primarily from forbidden lines have been able to
predict some aspects of observed spectra during late-nebular epochs, P-Cygni \ion{Fe}{2} absorption features and forbidden emission may be present during the first
couple hundred days post-maximum light.  In fact, the models in
Figure~\ref{fig:models} that best recreate the observed late-time spectra include both permitted
and forbidden lines \citep{Blo15}.  A few other models also show that the NIR
and optical features can be shaped, in part by, permitted lines
\citep{Fri14,Bra08,Bra09}.

\begin{figure}[t!]
        \centering
        \includegraphics[width=\columnwidth]{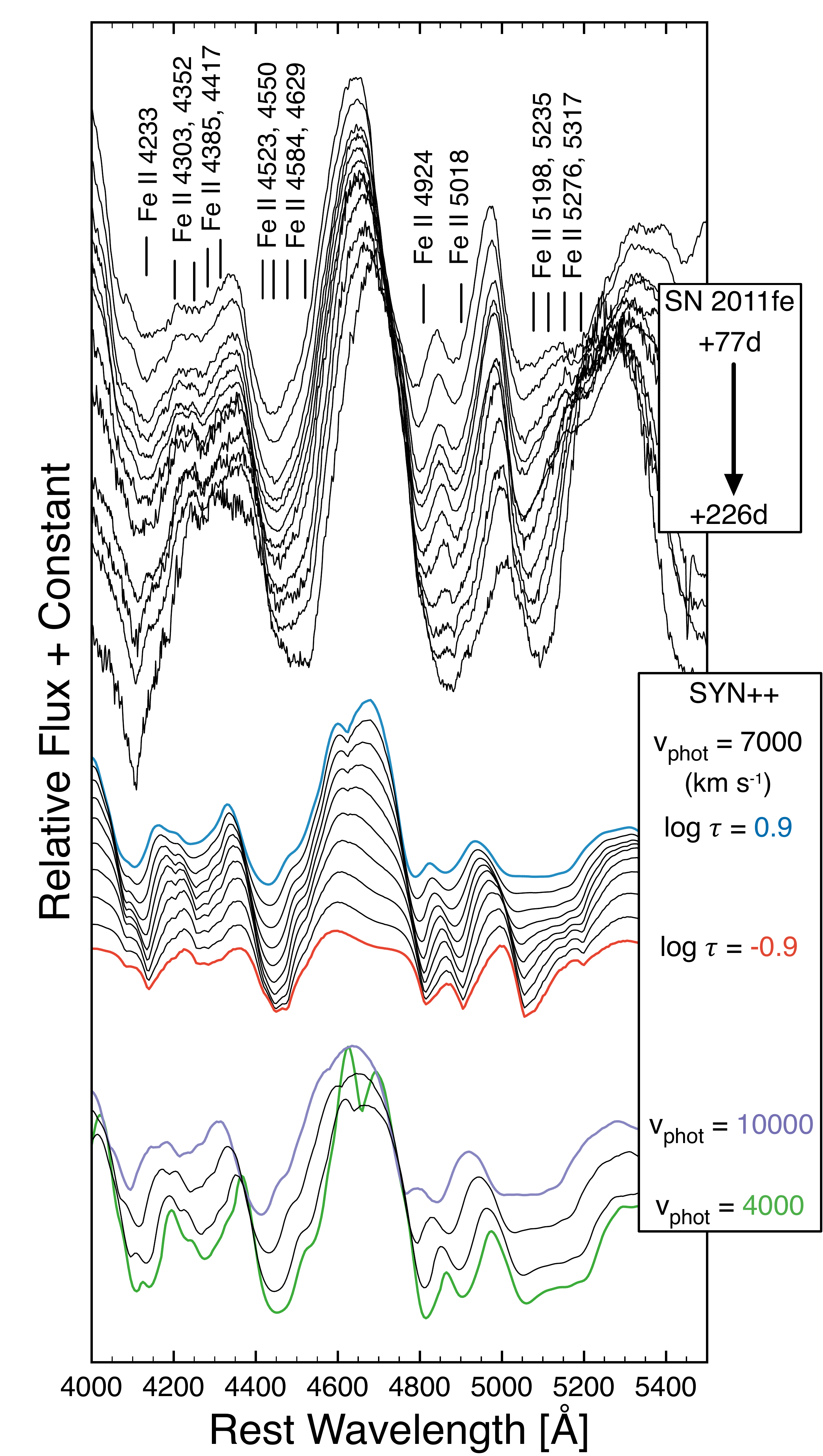}
        \caption{Spectra of SN 2011fe compared to two models computed using only
\ion{Fe}{2} with \texttt{SYN++}.  The first model maintains a constant photospheric
velocity of 7000~km~s$^{-1}$ and varies the opacity.  The second model varies
the velocity of the photosphere while successively decreasing the opacity from $log$ $\tau$ = 1.5 to 0 to
maintain relative line strengths.  Both models can be thought of as reproducing
the red-ward trend of the 4850 and 5000~\AA\ features. However, there is favorable evidence
that the spectral evolution can be significantly influenced by opacity effects
for permitted lines of \ion{Fe}{2}.}
\label{fig:synow} \end{figure}

To examine the presence of permitted lines as a means of
explaining some of the features and phenomena seen in late-time spectra, a
simple spectral model was computed using \texttt{SYN++} \citep{Tho11}.
Figure~\ref{fig:synow} shows the late-time spectra of SN 2011fe compared to two
toy models created with \texttt{SYN++} using only permitted \ion{Fe}{2} P-Cygni profiles. The first model, shown in the center of the figure, varied opacity
with time while maintaining a constant photospheric velocity of
7000~km~s$^{-1}$.  The second model, shown at the bottom of the figure, varied
the velocity of the photosphere from 10,000 km~s$^{-1}$ to 4000 km~s$^{-1}$
while successively increasing the opacity to maintain relative line strengths.

Since both \texttt{SYN++} models contain only permitted lines of \ion{Fe}{2},
we caution that these computed spectra are not meant to be representative of the
entire late-time evolution, as most features consist of blends, particularly for the 4700~\AA. Rather, this exercise 
shows that not only can the two minor features at 4850 and 5000~\AA\ be
reproduced with the use of only permitted lines, but that the red-ward motion of
these features can also be recreated through Doppler velocities and/or
decreasing opacities.

As suggested by previous works, line velocities can be invoked to explain the
red-ward motion of select features \citep{Mae10Jan,Mae11,Sil13}. Using \texttt{SYN++}, in
Figure~\ref{fig:synow} we show that a change in bulk line velocities of
\ion{Fe}{2}, starting near 10,000 km~s$^{-1}$ and evolving down to 4000
km~s$^{-1}$, can be used to follow the line centers for select features.
However, under an assumption of a sharp photosphere, decreasing opacity for a fixed expansion velocity can also cause most features, e.g., near 4850 and
5000~\AA, to shift to the red. 

By reducing the opacity over time, our simplified
\texttt{SYN++} model with a sharp photosphere can mimic a red-ward evolution for both 4850 and 5000~\AA\ features. Under the assumption of a transparent, continuum-emitting core, the emission peak of a profile is also predicted to move red-ward of the rest wavelength \citep{Fri12}. This suggests that the spectral evolution of some late-time features may be significantly influenced by declining opacities \citep{Bra08,Fri12}.

\section{Conclusions}

We have compiled a sample of 27 normal SN~Ia with a total of 160 spectra and
plotted the central wavelength of the 4700, 4850, 5000, 5500, and 5900~\AA\
features with respect to maximum brightness.  We propose
that the ubiquitous red shift of these common late-time, nebular SN~Ia spectral features
is not mainly due to a decrease in a blueshift of forbidden Fe lines as has
been proposed but the result, in part, of decreasing velocity and/or opacity of
P-Cygni Fe absorption. 
Specific findings of our study are the following:

(1) The late-time SN~Ia spectral features at 4700, 4850, and 5000 \AA\ all move to the red and do so at
roughly the same rate from day +50 out to at least day +300.  
Such progressive red shifts appear characteristic of all normal
Type Ia SNe and is key in understanding their correct identifications.

(2) The 4700~\AA\ feature shows an initial red shift of $\sim$0.3~\AA\ per day
at the beginning of the nebular phase, slowing to a rate of $\sim$0.2~\AA\ per
day.  This rate is maintained out to very late times (day 300+) with no
indication of slowing down, contrary to expectations if due to an initial
blueshift velocity of an [\ion{Fe}{3}] blend centered at 4701 \AA.  

(3) The initial rapid shift seen in the 4700 \AA\ feature is likely due to the decay of [\ion{Co}{2}] 4623 \AA\ and the strengthening of [\ion{Fe}{3}] 4658 \AA.  However, our models of permitted Fe II absorption suggest that [\ion{Fe}{3}] is not the main source for the 4700 \AA\ feature's intensity and central wavelength for all epochs during late-nebular phases.  This does not imply that [\ion{Fe}{3}] is absent from this feature, but rather that the contributions from [\ion{Fe}{3}], as well as possibly [\ion{Fe}{2}] and [\ion{Co}{2}], are weaker than previous models suggest.

(4) The FWHM of the 4700~\AA\ feature is found to increase significantly with time.
Part of this increase appears to be due to the emergence of additional flux on
its red side, and this increase may not be uncommon in SNe~Ia.  In contrast, however, the
widths of the 4850 and 5000 \AA\ features do not change until very late and
show no evidence of blending.

(5) The feature at 5900~\AA\ also shows a shift to the red, though
not at the same rate as the three bluer features at late times.  This feature
moves at roughly 0.4 $\pm0.2$~\AA\ per day or less, slowing to about 0.1~\AA\ after
day~$+$100.  We find that this feature exhibits a large range in central wavelengths across our SN sample at both early and late times.
\smallskip

\indent To better understand the evolution and composition of late-time SN~Ia spectra, including the relative abundances and expansion
velocities of the Fe-peak element rich ejecta, future theoretical
models must take into
account the observed red shifts of blue nebular features described here.  Illustrating the limits of our comprehension of late-time spectral
features are the large and currently mysterious red shifts of several optical
features reported in the very late-time spectrum of the bright SN~Ia event,
SN~2011fe, out to day $+$1000 \citep{Tau15,Gra15}.  Investigating the cause
behind the  phenomenon of late-time red shifting will hopefully lead to better
insights into the nature of the nebular phase of SNe Ia. 

\acknowledgments
The authors wish to thank the referee for providing useful comments and suggestions, M. Modjaz and P. Challis for obtaining the Keck and MMT data on
SN~2008Q.  We also wish to thank L. Dessert and S. Blondin for making public their model spectra and the staff of MDM Observatory for their assistance in making these observations possible.
CSB's supernova research is supported in part by a Fellowship from Dartmouth's School of Graduate and Advanced Studies. RAF's type Ia supernova research is funded in part by NASA/STScI grants 12609 and 13471.
This work was made possible by contributions to the Supernova Spectrum Archive \citep{Ric01}, the Weizmann Interactive Supernova data REPository (WISeREP \citealt{Yar12}), and the Berkeley SN Ia Program (BSNIP \citealt{Sil12}). Several model spectra, as well as some of the data that were not available on WISeREP at the time of our initial study were obtained with the help of the graph digitizer software, \textsc{graphclick}.\footnote{The full software is available at http://www.arizona-software.ch/graphclick/.}

\bibliographystyle{apj}
\bibliography{Bib_master}

\appendix

\LongTables
\begin{deluxetable}{lccccccc}[b]
  \tablecolumns{8}
  \tablewidth{\linewidth} 
  \tablecaption{Measured Central Wavelengths of Late-Time Sn Ia Features}
  \tablehead{
  \colhead{SN} & \colhead{Post-Max} & \colhead{4700 \AA}& \colhead{4700 \AA\ FWHM}& \colhead{4850 \AA} & \colhead{5000 \AA} & \colhead{5500 \AA}& \colhead{5900 \AA}\\
     & \colhead{Epoch} & \colhead{(\AA)}& \colhead{(\AA)}& \colhead{(\AA)} &\colhead{(\AA)} &\colhead{(\AA)}& \colhead{(\AA)}}
  \startdata
	1990N & 186 & 4673 & 172 &   & 4973 &   & 5895 \\ 
	 & 227 & 4685 & 190 &   & 4968 &   & 5898 \\ 
	 & 255 & 4689 & 187 &   & 4970 &   & 5900 \\ 
	 & 280 & 4693 & 181 &   & 4983 &   & 5901 \\ 
	 & 333 & 4699 & 217 &   &   &   &   \\ 
	1994D & 068 & 4631 & 149 & 4833 & 4959 & 5515 & 5878 \\ 
	 & 069 & 4631 & 152 & 4837 & 4957 & 5512 & 5875 \\ 
	 & 073 & 4630 & 156 & 4834 & 4959 &   & 5880 \\ 
	 & 088 & 4638 & 166 & 4838 & 4965 &   & 5896 \\ 
	 & 097 & 4645 & 176 & 4841 & 4968 &   & 5899 \\ 
	 & 101 & 4647 & 178 & 4842 & 4969 &   & 5902 \\ 
	 & 129 & 4662 & 183 &   & 4973 &   & 5910 \\ 
	1994ae & 078 & 4628 & 169 & 4833 & 4953 & 5515 & 5863 \\ 
	 & 084 & 4632 & 185 & 4833 & 4957 & 5515 & 5871 \\ 
	 & 102 & 4642 & 185 & 4838 & 4961 & 5509 & 5876 \\ 
	 & 159 & 4662 & 188 &   & 4979 &  & 5893 \\ 
	 & 168 & 4665 & 184 &   & 4980 & & 5890 \\ 
	 & 234 & 4684 & 190 &   &   &   \\ 
	 & 383 & 4710 & 210 &   &   &   \\ 
	1995D & 071 & 4625 & 153 & 4829 & 4955 & 5508 & 5855 \\ 
	 & 075 & 4626 & 155 & 4832 & 4957 & 5502 & 5858 \\ 
	 & 080 & 4632 & 160 & 4842 & 4957 & 5505 & 5861 \\ 
	 & 103 & 4643 & 175 & 4841 & 4964 &   & 5871 \\ 
	 & 287 & 4688 & 210 &   & 4972 &   &   \\ 
	 & 295 & 4693 & 213 &   &   &   &   \\ 
	1996X & 057 & 4640 & 136 & & 4971 & & 5891 \\ 
	 & 087 & 4660 & 165 & & 4985 & & 5914 \\ 
	 & 298 & 4717 & 170 & & 4982 & & 5924 \\ 
	1998aq & 060 & 4620 & 153 & 4824 & 4947 & 5496 & 5848 \\ 
	 & 063 & 4620 & 156 & 4824 & 4948 & 5495 & 5850 \\ 
	 & 066 & 4623 & 154 & 4825 & 4950 & 5494 & 5854 \\ 
	 & 079 & 4631 & 176 & 4829 & 4954 & 5490 & 5861 \\ 
	 & 082 & 4634 & 177 & 4828 & 4956 & 5490 & 5860 \\ 
	 & 091 & 4636 & 180 & 4834 & 4955 &   & 5865 \\ 
	 & 211 & 4665 & 201 &   &   \\ 
	 & 231 & 4675 & 198 &   & 4987 &   & 5895 \\ 
	 & 241 & 4677 & 202 &   & 4986 &   &  5900 \\ 
	1998bu & 058 & 4624 & 162 & 4826 & 4949 & 5506 & 5855\\ 
	 & 178 & 4669 & 191 &   & 4977 &   & 5883\\ 
	 & 189 & 4670 & 188 &   & 4976 &   & 5885 \\ 
	 & 207 & 4674 & 191 &   & 4974 &   & 5885 \\ 
	 & 216 & 4676 & 191 &   & 4972 &   & 5882 \\ 
	 & 235 & 4679 & 191 &   & 4966 &   & 5889\\ 
	 & 242 & 4678 & 195 &   & 4959 &   & 5890 \\ 
	 & 279 & 4685 & 195 &   & 4952 &   & 5905\\ 
	 & 328 & 4697 & 206 &   & 4951 &   & 5903\\ 
	2002bo & 057 & 4652 & 158 &   & 4979 & 5570 & 5904 \\ 
	 & 073 & 4662 & 184 &   & 4988 & 5567 & 5917 \\ 
	 & 080 & 4667 & 185 &   & 4992 & 5564 & 5923 \\ 
	2002dj & 082 & 4650 & 187 &   & 4960 & 5547 & 5890 \\ 
	 & 234 & 4692 & 189 &   & 5000 &   & 5930 \\ 
	2002dj & 287 & 4695 & 197 &   & 4980 &   & 5915 \\ 
	2003du & 063 & 4626 & 152 & 4830 & 4956 & 5517 & 5865 \\ 
	 & 072 & 4629 & 158 & 4829 & 4958 & 5514 & 5869 \\ 
	 & 084 & 4638 & 173 & 4834 & 4960 & 5515 & 5875 \\ 
	 & 087 & 4639 & 170 & 4843 & 4959 & 5514 & 5877 \\ 
	 & 109 & 4654 & 187 &   & 4966 &   & 5886 \\ 
	 & 138 & 4658 & 189 &   & 4971 &   & 5885 \\ 
	 & 139 & 4660 & 188 &   & 4971 &   & 5886\\ 
	 & 142 & 4662 & 186 &   & 4973 &   & 5889 \\ 
	 & 196 & 4675 & 180 & & 4985 \\ 
	 & 209 & 4681 & 175 &   & 4989 &   \\ 
	 & 221 & 4682 & 188 &   & 4975 &   & 5889 \\ 
	 & 272 & 4686 & 205 &   &   &   &   \\ 
	 & 377 & 4710 & 213 &   &   \\ 
	2003hv & 071 & 4617 & 151 & 4835 & 4953 & & 5865 \\ 
	 & 073 & 4618 & 155 & 4834 & 4952 &   & 5864 \\ 
	 & 078 & 4623 & 163 & 4834 & 4954 &   & 5866 \\ 
	 & 108 & 4636 & 174 &   & 4961 &   & 5873 \\ 
	 & 140 & 4651 & 178 &   & 4963 &   & 5874 \\ 
	 & 318 & 4689 & 195 &   & 4933 &   &   \\ 
	2003kf & 063 & 4620 & 173 & 4824 & 4948 & 5524 & 5860 \\ 
	 & 078 & 4634 & 177 & 4824 & 4955 & 5522 & 5875 \\ 
	 & 080 & 4634 & 185 & 4823 & 4953 & 5524 & 5876 \\ 
	 & 111 & 4654 & 189 &   & 4965 & 5523 & 5890 \\ 
	2004dt & 061 & 4626 & 178 & 4840 & 4959 & 5508 & 5844 \\ 
	 & 063 & 4628 & 178 & 4843 & 4962 & 5507 & 5841 \\ 
	 & 065 & 4630 & 175 & 4850 & 4962 & 5505 & 5846 \\ 
	 & 095 & 4649 & 191 & 4852 & 4975 & 5496 & 5847 \\ 
	2004eo & 057 & 4619 & 155 & 4824 & 4948 & 5522 & 5857 \\ 
	 & 071 & 4630 & 157 & 4830 & 4960 & 5522 & 5866 \\ 
	 & 075 & 4625 & 154 & 4835 & 4956 & 5514 & 5872 \\ 
	 & 231 & 4674 & 166 &   & 4978 &   & 5887 \\ 
	2005cf & 061 & 4622 & 167 & 4825 & 4955 & 5524 & 5863 \\ 
	 & 075 & 4629 & 162 & 4839 & 4958 & 5519 & 5873 \\ 
	 & 077 & 4635 & 161 & 4833 & 4962 & 5528 & 5877 \\ 
	 & 083 & 4638 & 178 & 4844 & 4963 & 5528 & 5883 \\ 
	 & 319 & 4689 & 197 &   &   &   &   \\ 
	2007af & 091 & 4646 & 157 & 4852 & 4972 & 5529 & 5894 \\ 
	 & 113 & 4656 & 165 & 4860 & 4980 &   & 5899 \\ 
	 & 121 & 4659 & 163 &   & 4984 &   & 5896 \\ 
	 & 144 & 4662 & 170 &   & 4983 &   & 5906 \\ 
	 & 158 & 4668 & 185 &   & 4991 &   & 5900 \\ 
	2007gi & 086 & 4651 & 186 &   & 5007 &   & 5911 \\ 
	 & 150 & 4677 & 188 &   & 5005 &   & 5924 \\ 
	2007le & 066 & 4635 & 181 &   & 4946 & 5589 & 5885 \\ 
	 & 077 & 4645 & 186 &   & 4963 & 5553 & 5896 \\ 
	 & 305 & 4690 & 194 &   & 4967 &   &  5904 \\ 
	2007sr & 071 & 4648 & 180 &   & 4959 & 5549 & 5896 \\ 
	 & 079 & 4651 & 178 &   & 4966 & 5547 & 5901 \\ 
	 & 103 & 4662 & 176 &   & 4978 &   & 5911 \\ 
	 & 111 & 4664 & 179 &   &   &   & 5913 \\ 
	 & 129 & 4668 & 171 &   &   &   & 5913 \\ 
	 & 185 & 4680 & 179 &   &   &   & 5914 \\ 
	2008Q\,\tablenotemark{a} & 203 & 4676 & 197 & & 4961 & & 5897 \\ 
	 & 237 & 4694 & 202 & & 4967 & & 5909 \\ 
	 & 318 & 4698 & 222 &   &   \\ 
	2011by & 051 & 4631 & 144 & 4838 & 4962 & 5527 & 5864 \\ 
	 & 206 & 4686 & 179 &   & 4994 &   & 5900 \\ 
	 & 310 & 4705 & 187 &   &   &   & 5901 \\ 
	2011fe & 074 & 4636 & 153 & 4841 & 4966 & 5520 & 5878 \\ 
	 & 077 & 4637 & 156 & 4843 & 4968 & 5519 & 5879 \\ 
	 & 079 & 4639 & 155 & 4845 & 4970 & 5521 & 5881 \\ 
	 & 082 & 4640 & 161 & 4845 & 4973 & 5520 & 5881 \\ 
	 & 084 & 4641 & 158 & 4844 & 4973 & 5518 & 5884 \\ 
	 & 087 & 4641 & 162 & 4846 & 4974 & 5520 & 5885 \\ 
	 & 089 & 4642 & 165 & 4846 & 4976 & 5518 & 5886 \\ 
	 & 100 & 4648 & 167 & 4849 & 4979 & 5517 & 5890 \\ 
	 & 103 & 4648 & 166 & 4850 & 4979 &   & 5889 \\ 
	 & 107 & 4650 & 168 & 4853 & 4980 &   \\ 
	 & 115 & 4655 & 169 & 4853 & 4983 &   & 5894 \\ 
	 & 130 & 4657 & 170 & 4858 & 4986 &   & 5896 \\ 
	 & 145 & 4664 & 168 & 4860 & 4990 &   & 5898 \\ 
	 & 166 & 4668 & 177 & 4866 & 4993 &   & 5901 \\ 
	 & 174 & 4672 & 175 &   & 4996 &   & 5899 \\ 
	 & 226 & 4685 & 176 &   & 5005 &   & 5909 \\ 
	2011fe & 232 & 4686 & 178 &   & 4997 &   \\ 
	 & 259 & 4691 & 178 &   & 5000 &   & 5910 \\ 
	 & 289 & 4695 & 179 &   & 4997 &   & 5910 \\ 
	 & 346 & 4703 & 191 &   & 4979 &   & 5916 \\ 
	 & 348 & 4703 & 192 &   & 4975 &   & 5916 \\ 
	2012fr & 062 & 4629 & 200 &   & 4957 & 5556 & 5878 \\ 
	 & 071 & 4636 & 215 &   & 4928 & 5562 & 5894 \\ 
	 & 077 & 4647 & 208 &   & 4929 & 5562 & 5899 \\ 
	 & 118 & 4665 & 190 &   & 4919 & 5561 & 5914 \\ 
	 & 151 & 4674 & 182 &   & 4926 &   & 5925 \\ 
	 & 222 & 4682 & 197 &   & 5011 &   & 5919 \\ 
	 & 261 & 4689 & 201 &   & 4991 &   & 5912 \\ 
	 & 340 & 4689 & 215 &   &   &   &  5910\\ 
	 & 367 & 4689 & 216 &   &   &   &   \\ 
	2012hr & 283 & 4686 & 171 &   & 4971 & & 5904 \\ 
	 & 368 & 4696 & 203 &   &   &   \\ 
	2013aa & 137 & 4661 & 181 &   & 4977 &   & 5882 \\ 
	 & 185 & 4672 & 181 &   & 4982 &   & 5884 \\ 
	 & 205 & 4677 & 180 &   & 4984 &   & 5886 \\ 
	 & 345 & 4701 & 205 &   &   &   &   \\ 
	2013dy & 071 & 4654 & 182 & 4850 & 4976 & 5531 & 5888 \\ 
	 & 076 & 4658 & 183 & 4847 & 4974 & 5530 & 5890 \\ 
	 & 092 & 4666 & 185 & 4846 & 4981 & 5520 & 5898 \\ 
	 & 100 & 4668 & 189 &   & 4984 & 5526 & 5902 \\ 
	 & 125 & 4675 & 192 &   & 4989 &   & 5905 \\ 
	 & 130 & 4676 & 187 &   & 4991 &   & 5905 \\ 
	 & 133 & 4676 & 190 &   & 4990 &   & 5908 \\ 
	 & 333 & 4703 & 215 \\ 
	 & 423 & 4728 & 233 \\ 
	2014lp & 087 & 4647 & 175 & & 4958 & 5000 & 5543 \\ 
	 & 114 & 4660 & 175 &  & 5005 &   \\ 
	 & 149 & 4668 & 178 &  & 5016 & 
   \enddata
    \tablecomments{Errors for the measured central wavelengths and FWHM are $\pm$4 \AA\ and $\pm$5 \AA\, respectively.}
    \tablenotetext{a}{Our day +237 and +318 spectra of SN2008Q are available on WISeREP.}
\end{deluxetable}

\end{document}